\documentclass[fleqn,11pt]{article}
\usepackage{amsmath,amssymb}
\usepackage{xcolor}
\usepackage{verbatim}

\newtheorem{stat}{Statement}

\newtheorem{Mark}{Comment}
\newcommand{\Marking}[2]{%
\begin{Mark}\hspace{-4pt}\textbf{(#1).}\\
#2
\end{Mark}}

\newcommand{\Mod}[1]{\mathbf{#1}}
\newcommand{\ModTilde}[1]{\tilde{\mathbf{#1}}}
\newcommand{\Fig}[3]{%
\begin{center}
\parbox{8cm}{%
\refstepcounter{figure}\includegraphics[width=8cm,height=#2cm]{#1} \noindent Fig. \thefigure:\quad
#3}\end{center}}
\newcounter{strochka}

\newcounter{spisok}
\usepackage{amsfonts,amssymb,cite}
\usepackage{graphicx}

\def\noi{\noindent}

\newcommand{\Title}[1]{\noi {{\Large\bf #1}}\\[1ex]}

\newcommand{\Author}[2]{\noi{\bf #1}\\[2ex]\noi{\normalsize\it #2}\\}

\newcommand{\Abstract}[1]{\vskip 2mm \begin{center}
        \parbox{16.4cm}{\small\noi #1} \end{center}\medskip}
\newcommand{\foom}[1]{\protect\footnotemark[#1]}

\def\nqq{\hspace*{-2em}}

\def\Jl#1#2{#1 {\bf #2},\ }

\def\ApJ#1 {\Jl{Astroph. J.}{#1}}
\def\CQG#1 {\Jl{Class. Quantum Grav.}{#1}}
\def\DAN#1 {\Jl{Dokl. AN SSSR}{#1}}
\def\GC#1 {\Jl{Grav. Cosmol.}{#1}}
\def\GRG#1 {\Jl{Gen. Rel. Grav.}{#1}}
\def\JETF#1 {\Jl{Zh. Eksp. Teor. Fiz.}{#1}}
\def\JETP#1 {\Jl{Sov. Phys. JETP}{#1}}
\def\JHEP#1 {\Jl{JHEP}{#1}}
\def\JMP#1 {\Jl{J. Math. Phys.}{#1}}
\def\NPB#1 {\Jl{Nucl. Phys. B}{#1}}
\def\NP#1 {\Jl{Nucl. Phys.}{#1}}
\def\PLA#1 {\Jl{Phys. Lett. A}{#1}}
\def\PLB#1 {\Jl{Phys. Lett. B}{#1}}
\def\PRD#1 {\Jl{Phys. Rev. D}{#1}}
\def\PRL#1 {\Jl{Phys. Rev. Lett.}{#1}}

\def\lal{&&\nqq {}}

\def\beq{\begin{equation}}
\def\eeq{\end{equation}}
\def\bear{\begin{eqnarray}}
\def\bearr{\begin{eqnarray} \lal}
\def\ear{\end{eqnarray}}
\def\earn{\nonumber \end{eqnarray}}

\begin{document}
\thispagestyle{empty}
\twocolumn[

\vspace{1cm}

\Title{Formation of supermassive nuclei of Black holes in the early\\[8pt] Universe by the mechanism of scalar-gravitational instability.\\[8pt] II. The evolution of localized spherical perturbations.  \foom 1}

\Author{Yu. G. Ignat'ev}
    {Institute of Physics, Kazan Federal University, Kremlyovskaya str., 16A, Kazan, 420008, Russia}


\Abstract
 {Based on the previously formulated theory of spherical perturbations in the cosmological medium of self-gravitating scalarly charged fermions with the Higgs scalar interaction and the similarity properties of such models, the formation of supermassive Black Hole Seeds (SSBH) in the early Universe is studied. Using numerical simulation of the process, it is shown that the mass of SSBH during the evolution process reaches a limiting value, after which it begins to slowly fall. The possible influence of nonlinearity on this process is discussed.
}
\bigskip

] 

\section{Introduction}
In the first part of the author's article \cite{Yu_GC_23_No4} two problems were formulated that need to be solved in the theory of the formation of supermassive black hole (SSBH) nuclei in the early Universe using the mechanism of scalar-gravitational instability of the cosmological medium of scalar-charged fermions, in order to lead it in accordance with the observed picture:\\
1. According to the results presented, the process of increasing the SSBH mass does not stop when the required mass \eqref{M_nc} is reached (see \cite{SMBH1e} -- \cite{SMBH2e})
\begin{equation}\label{M_nc}
m_{ssbh}\sim 10^4\div 10^6 M_\odot\approx 10^{42}\div10^{44}m_{\mathrm{pl}},
\end{equation}
but continues endlessly. Now we need to find a mechanism to stop this process.\\
2. What does the large-scale structure of the Universe become after the completion of this process, what is the fate of the matter that fell into the sphere of influence of SSBH?

In this part of the work we will try to solve the first of these problems, referring the solution to the second in the third part of the article.

\section{Evolution of small localized\newline spherical perturbations in\newline a medium of scalarly charged degenerate fermions}
To solve the first of these problems, let us consider the evolution of small localized spherical perturbations in the cosmological medium of scalarly charged degenerate fermions \cite{YuTMF_23}.
\subsection {Small spherical perturbations}
We write the metric with gravitational perturbations in isotropic spherical coordinates with a conformally Euclidean metric of three - dimensional space\footnote{see, for example, \cite{Land_Field}}, al\-lo\-wing a continuous transition to the Friedmann metric:
\begin{eqnarray}
\label{metric_pert}
ds^2=\mathrm{e}^{\nu(r,t)}dt^2-a^2(t)\mathrm{e}^{-\nu(r,t)}[dr^2\nonumber\\
+r^2(d\theta^2+\sin^2\theta d\varphi^2)],
\end{eqnarray}
where $\nu(r,t)$ and $\lambda(r,t)$ are small longitudinal perturbations of the Friedmann metric ($\nu\ll1,\ \lambda\ll1$). Let us introduce small perturbations of fermions and the scalar field\footnote{For notation, see the first part of the work\cite{Yu_GC_23_No4}.}
\begin{eqnarray}\label{dF-drho-du}
\begin{array}{lcl}
\Phi(r,t)&=&\Phi(t)+\varphi(r,t);\\
\pi_z(r,t)&=&\pi_z(t)(1+\delta_z(r,t));\\
\sigma(r,t)&=& \sigma(t)+\delta\sigma(r,t);\\
u^i&=&\displaystyle \delta^i_4(1-\nu(r,t))+\delta^i_1 v(r,t).\\
\end{array}
\end{eqnarray}

According to \cite{YuPhysA}, let us select the particle-like singular part of the solution of the perturbed Einstein equations and the scalar Higgs field with a source, introducing the scalar functions $\rho(r,t)$ and $\chi(r,t)$, which are nonsingular at the origin of coordinates
\begin{eqnarray}\label{nu=1-2}
\nu(r,t)=2\frac{\rho(r,t)-m(t)}{a(t)r}\equiv -\frac{2\mu(r,t)}{a(t)r},\\
\label{phi=1-2}
\varphi(r,t)=2\frac{\chi(r,t)-q(t)}{a(t)r}\equiv \frac{2\phi(r,t)}{a(t)r},
\end{eqnarray}
where $m(t)$ and $q(t)$ are singular mass and scalar charge such that
\begin{eqnarray}\label{rho/r}
\rho(0,t)=0;\quad  \lim_{r\to 0}r\frac{\partial\rho(r,t)}{\partial r}=0,\\
\label{phi/r}
\chi(0,t)=0;\; \lim_{r\to 0}r\frac{\partial \chi(r,t)}{\partial r}=0.
\end{eqnarray}

In \cite{YuTMF_23} it is proved that the entire system of equations for disturbances can be written as a system of four linear homogeneous equations, of which the first two equations form an independent subsystem of ordinary homogeneous differential equations for the source functions $m(t)$ and $q(t )$ (\eqref{EQ_m} -- \eqref{EQ_q})
\begin{eqnarray}\label{EQ_m}
-3\frac{(1+\psi^2)}{\psi^2}\ddot{m}-3H\frac{2+3\psi^2}{\psi^2}\dot{m} &\nonumber\\
-Z\frac{3+2\psi^2}{\psi^2}\dot{q}+\mathrm{L}_{\varepsilon m}(t)m+\mathrm{L}_{\varepsilon q}(t)q =0;&\\
\label{EQ_q}
\ddot{q}+\biggl(H-\frac{3}{2\Phi \psi^2}\biggr)\dot{q} -\frac{3}{2\Phi\psi^2}\ddot{m}&\nonumber\\
+\biggl(\frac{3H}{\Phi\psi^2}-2Z\biggr)\dot{m}+\mathrm{L}_{\sigma m}(t)m+\mathrm{L}_{\sigma q}(t)q =0.&
\end{eqnarray}
and the remaining two equations constitute an independent subsystem of homogeneous partial\newline differential equations with respect to the functions $\rho(r,t)$ and $\chi(r,t)$ (\eqref{EQ_rho} -- \eqref{EQ_chi}) satisfying the condi\-tions \eqref{rho/r} and \eqref{phi/r} at the origin
\begin{eqnarray}\label{EQ_rho}
\frac{1}{a^2}\rho''-3\frac{(1+\psi^2)}{\psi^2}\ddot{\rho}-3H\frac{2+3\psi^2}{\psi^2}\dot{\rho}&\nonumber\\
-Z\frac{3+2\psi^2}{\psi^2}\dot{\chi}+\mathrm{L}_{\varepsilon m}(t)\rho+\mathrm{L}_{\varepsilon q}(t)\chi &=0;\\
\label{EQ_chi}
\ddot{\chi}+\biggl(H-\frac{3}{2\Phi\psi^2}\biggr)\dot{\chi}-\frac{1}{a^2}\chi'' -\frac{3}{2\Phi\psi^2}\ddot{\rho}&\nonumber\\
+\biggl(\frac{3H}{\Phi\psi^2}-2Z\biggr)\dot{\rho}+\mathrm{L}_{\sigma m}(t)\rho+\mathrm{L}_{\sigma q}(t)\chi &=0,
\end{eqnarray}
where the functions $\mathrm{L}_{\alpha\beta}(t)$ are introduced, defined by the background solutions $[\xi(t)$, $H(t)$, $\Phi(t)$, $Z (t)]$:
\begin{eqnarray}\label{L_em}
\mathrm{L}_{\varepsilon m}(t)=\frac{Z^2}{2}+\frac{4}{\pi}e^4\Phi^4\frac{(1+\psi^2)^{3/2}}{\psi};\nonumber\\
\label{L_eq}
\mathrm{L}_{\varepsilon q}(t)=HZ\frac{3+2\psi^2}{\psi^2}+\frac{3+4\psi^2}{\psi^2}\Phi(m_0^2-\alpha\Phi^2)\nonumber\\
+4e^4\Phi^3\frac{1+2\psi^2}{\pi\psi^2}F_1(\psi);\nonumber\\
%
\label{L_sm}
\mathrm{L}_{\sigma m}(t)=-2HZ-\Phi(m_0^2-\alpha\Phi^2)\nonumber\\
+\frac{2e^4\Phi^3}{\pi\psi}\bigl(\sqrt{1+\psi^2}-2\psi F_1(\psi)\bigr);\nonumber\\
\label{L_sq}
\mathrm{L}_{\sigma q}(t)=\frac{3HZ}{2\Phi\psi^2}-\frac{2}{3}\Lambda+\frac{3}{2\psi^2}(m_0^2-\alpha\Phi^2)\nonumber\\
+\frac{1}{6\alpha}(m_0^2-\Phi^2)^2)+\frac{Z^2}{6}+\frac{8e^4\Phi^2\psi^3}{\pi\sqrt{1+\psi^2}}\nonumber\\
+\frac{2e^4\Phi^2}{\pi\psi^2}F_1(\psi)\biggl(1-6\psi^2-\frac{\Phi^2\psi^2}{3}\biggr),
\nonumber
\end{eqnarray}
as well as kinetic functions $\psi,F_1(\psi),F_2(\psi)$
\begin{eqnarray}\label{psi(eta)}
\psi=\frac{\pi^0_z}{|e\Phi|}\mathrm{e}^{-\xi}, \quad (\pi^0_z=\pi_z(0));\nonumber\\
\label{F_1}
F_1(\psi)=\psi\sqrt{1+\psi^2}-\ln(\psi+\sqrt{1+\psi^2});\nonumber\\
\label{F_2}
\!\!\!F_2(\psi)=\psi\sqrt{1+\psi^2}(1+2\psi^2)-\ln(\psi+\sqrt{1+\psi^2}).\nonumber
\end{eqnarray}

We will call the equations \eqref{EQ_m} and \eqref{EQ_q} evolutionary equations for the mass and scalar charge of the singular source. The initial conditions for these equations are:
\begin{eqnarray}\label{Inits_pert}
m(t_0)=m_0,\; \dot{m}(0)=\dot{m}_0;\nonumber\\
q(0)=q_0,\;\dot{q}(0)=\dot{q}_0.
\end{eqnarray}
Since these equations are linear and homogeneous, then under zero initial conditions
$m_0=0, \dot{m}_0=0$; $q_0=0,\dot{q}_0=0$they only have a trivial solution$m(t)=0,\ q(t)=0$.

If at least one of the initial conditions is not zero, then $m(t)\not=0,\ q(t)\not=0$, which is necessary for the formation of a scalar black hole.
Although the two subsystems of equations \eqref{EQ_m} - \eqref{EQ_q} and \eqref{EQ_rho} and \eqref{EQ_chi} are independent, their solutions can be related by the boundary conditions of the Cauchy problem.

\subsection{Localized Spherical Perturbations}
To solve the Cauchy problem inside a sphere of radius $r_0$, it is necessary to specify perturbations inside this sphere at the zero moment of time\footnote{The initial time $t_0=0$ was chosen for convenience.}:
\begin{equation}\label{Coshe}
\rho(r,0)=\rho(r);\qquad \phi(r,0)=\phi(r),
\end{equation}
where $\rho(r)$ and $\chi(r)$ are given functions.

The concept of localized spherical perturbations was introduced in the works of \cite{YuPhysA} (for more details, see \cite{YuTMF_23}). Localization of spherical disturbances at the initial moment of time means that
at this moment of time $t_0=0$ at some radius $r_0$ in the metric \eqref{metric_pert} zero boundary conditions are satisfied for the perturbations $\nu(r,t)$, $\varphi(r,t)$ and their first derivatives with respect to radius. With respect to the nonsingular scalar functions $\rho(r,t)$ and $\chi(r,t)$, these conditions can be rewritten as:
\begin{eqnarray}\label{bound_1}
\rho(r_0,0)=\mu(0); \quad {\displaystyle \left.\frac{d\rho(r,0)}{dr}\right|_{r=r_0}=0}; \nonumber\\
\chi(r_0,0)=q(0); \quad {\displaystyle  \left.\frac{d\chi(r,0)}{dr}\right|_{r=r_0}=0}.
\end{eqnarray}

According to \eqref{nu=1-2} and \eqref{phi=1-2} near the singularity $r\to0$ the perturbed metric \eqref{metric_pert} in an approximation linear in the smallness of the perturbations can be written in the form of a Schwarzschild metric with a physical radius $ R=ar$
{\small
\begin{eqnarray}\label{shvarc}
\!\!\!\!\!ds^2\backsimeq\!  \biggl(1\!-\!\frac{2m}{a(t)r}\biggr)dt^2 \!\!
  -\frac{dr^2+r^2(d\theta^2+\sin^2\theta d\varphi^2)}{\displaystyle1-\frac{2m}{ar}}.
\end{eqnarray}
}
The localization of spherical disturbances\\ means that inside the localization sphere the pertur\-ba\-tions can be represented as a Taylor series in powers of the radius, which, in turn, allows the system of equations \eqref{EQ_rho} -- \eqref{EQ_chi} for the functions $\rho(r ,t)$ and $\chi(r,t)$ are represented as a chain of ordinary interlocking differential equations. If at the initial moment of time the perturbations are representable in the form of finite series, these chains can be looped using the boundary conditions \eqref{bound_1} and show, firstly, that the localization radius of perturbations in isotropic spherical coordinates does not depend on time, i.e. That is, the physical radius of localization of the perturbations is pro\-por\-ti\-onal to the scale factor:
\begin{equation}\label{r(t)}
r_m=a(t)r_0.
\end{equation}
In this case, the total mass of the perturbations inside the localization sphere, $\delta M(r_0)$, is strictly equal to zero
\begin{equation}\label{dM}
\delta M(r_0,t)=0,
\end{equation}
which is a consequence of Birkhoff's theorem, i.e., the law of conservation of energy.

Secondly, the chain of equations obtained in \cite{YuTMF_23} for the coefficients $\rho_k(t),\ \chi_k(t)$ of the expansion of nonsingular parts of perturbations $\rho(r,t),\ \chi(r,t) $ through boundary conditions \eqref{bound_1} becomes inho\-mo\-ge\-neous, -- their right-hand sides are determined by solutions of the system of homogeneous equations \eqref{EQ_m} -- \eqref{EQ_q} with respect to $m(t),q(t)$, defining the singular part perturbations. Note that, thus, the evolution of the singular mass and charge does not depend on the type of perturbation inside the localization radius, determined by the non-singular parts of the perturbations\footnote{As long as the perturbations inside the localization radius remain small and the evolution of the singular mass and charge is determined by the unperturbed cosmological solution $ \xi(t),H(t),\Phi(t),Z(t)$.}. While we are not interested in the details of the evolution of the perturbations within the radius of its localization, but only in the evolution of the central mass and scalar charge of the black hole, we can limit ourselves to solving these two equations.

\subsection{Similarity properties of a\newline mathematical model}
We use methods of similarity theory of mathematical models developed in \cite{Yu_Similar}, expanding them to a per\-turbed cosmological model. These methods are all the more important in this work, since technical difficulties
numerical integration of a unified system of nonlinear differential equations describing the background cosmological solution $[\xi(t)$, $H(t)$, $\Phi(t)$, $Z(t)]$\footnote{See. the first part of the work \cite{Yu_GC_23_No4}} and linear relative to singular mass $m(t)$ and charge $q(t)$ differential equations \eqref{EQ_m} -- \eqref{EQ_q} with coefficients determined by background solutions (t i.e., again, essentially nonlinear equations), do not allow free manipulation of the model parameters to study their influence on its properties.

Let us consider the similarity transformation of the model of self-gravitating scalarly charged fermions\cite{Yu_Similar}:
\begin{eqnarray}\label{trans_param}
\mathcal{S}_k(\Mod{M}): &  \alpha=k^2\tilde{\alpha},\; m_s=k\tilde{m}_s;& \nonumber\\
& e=\sqrt{k}\tilde{e};\;\Lambda= k^2\tilde\Lambda;&\\
\label{trans_x}
& x^i= k^{-1}\tilde{x}^i,\quad \pi_f= \sqrt{k}\tilde{\pi}_f, & .
\end{eqnarray}

Let's consider two cosmological models: $\mathbf{M}$ with fundamental parameters $\mathbf{P}$ and initial conditions $\mathbf{I}$
and a similar model $\tilde{\mathbf{M}}$ with fundamental parameters $\tilde{\mathbf{P}}$ and initial conditions $\tilde{\mathbf{I}}$ --
\begin{eqnarray}
\label{Inits_tilde}
\tilde{\mathbf{I}}=\biggl[\Phi_0,\frac{1}{k}Z_0\biggl];\\
\label{Par_tilde}
\tilde{\mathbf{P}}=\biggl[\biggl[\frac{\alpha}{k^2},\frac{m_s}{k},\frac{e}{\sqrt{k}},\frac{\pi_0}{\sqrt{k}}\biggr],\frac{\Lambda}{k^2}\biggr].
\end{eqnarray}

Table \ref{Similar_table} shows the rules for transforming the similarity of the main quantities of the model studied here.
\begin{center}
\refstepcounter{table}
Table \thetable. Transformation of similarity of the main quantities of the model.\\[6pt]
\label{Similar_table}
\begin{tabular}{l|r}
\hline
\parbox{3.8cm}{\vspace{2pt}System $\Mod{M}$\vspace{2pt}} & \parbox{3.8cm}{\vspace{2pt}System $\ModTilde{M}=S_k(\Mod{M})$\vspace{2pt}}\\
\hline
\parbox{3.8cm}{\vspace{2pt}Constants: $[[\alpha,m,e,\pi_0],\Lambda]$\vspace{2pt}} & \parbox{3.8cm}{\vspace{2pt}Constants: $\bigl[\bigl[\frac{\alpha}{k^2},\frac{m}{k},\frac{e}{\sqrt{k}},\frac{\pi_0}{\sqrt{k}}\bigr],\frac{\Lambda}{k^2}\bigr]$ \vspace{2pt}}\\
\hline
\parbox{3.8cm}{\vspace{2pt}Basic Variables\\[6pt] $[[\xi(t),H(t),\Phi(t)$,\\[6pt]
  $Z(t)],t,n],$\vspace{2pt}} & \parbox{3.8cm}{\vspace{2pt}Basic Variables\\[6pt]
  $\bigl[\bigl[\xi\bigl(\frac{t}{k}\bigr),\frac{1}{k}H\bigl(\frac{t}{k}\bigr),\Phi\bigl(\frac{t}{k}\bigr)$,\\[6pt]
  $\frac{1}{k}Z\bigl(\frac{t}{k}\bigr)\bigr],kt,\frac{n}{k}\bigr]$ \vspace{2pt}}\\
\hline
\parbox{3.8cm}{\vspace{2pt}Total sphere energy\\[6pt]
 $M_n(t)=4\pi H^2(t)\frac{a^3(t)}{n^3}$\vspace{2pt}} & \parbox{3.8cm}{\vspace{2pt $\tilde{M}_{\tilde{n}}(t)=kM_n\bigl(\frac{t}{k}\bigr)$\vspace{2pt}}}\\
\hline
\parbox{3.8cm}{\vspace{2pt}Perturbations\\
 $\nu(r,t),\varphi(r,t)$,\\ $\rho(r,t),\chi(r,t), m(t),q(t)$}
  & \parbox{3.8cm}{\vspace{2pt $k\tilde{\nu}\bigl(\frac{r}{k},\frac{t}{k}\bigr),\ k\tilde{\varphi}\bigl(\frac{r}{k},\frac{t}{k}\bigr)$\\
 $k\rho\bigl(\frac{t}{k}\bigr),k\chi\bigl(\frac{t}{k}\bigr)$,\\ $km\bigl(\frac{t}{k}\bigr), kq\bigl(\frac{t}{k}\bigr)$ }}\\
\hline
\end{tabular}
\end{center}

\Marking{to Table \ref{Similar_table}\label{mark_table}}{
According to the general properties of similarity, the quantities of interest to us $m(t)$ and $q(t)$ - the total mass and scalar charge - the charge of the black hole are transformed according to the formulas of the Table \ref{Similar_table}. But at the same time, the initial conditions for mass and charge \eqref{Inits_pert} must also be subject to a similarity transformation%
\begin{eqnarray}\label{Inits_pert_tilde}
\tilde{m}(t_0)=km_0\biggl(\frac{t_0}{k}\biggr),& \tilde{\dot{m}}(t_0)=k^2\dot{m}_0\biggl(\frac{t_0}{k}\biggr);\nonumber\\
\tilde{q}(t_0)=kq_0\biggl(\frac{t_0}{k}\biggr),&\tilde{\dot{q}}(t_0)=k^2\dot{q}_0\biggl(\frac{t_0}{k}\biggr).
\end{eqnarray}
Due to the linearity and homogeneity of the evolution equations \eqref{EQ_m} - \eqref{EQ_q} their solutions are obviously proportional to the values of the initial quantities. Therefore, if the initial conditions for such a model are preserved, the total mass and charge are preserved in such a model.
}
\subsection{Singular points of a dynamical system}
The following property of the $\Mod{M}$ model (dynamical system) holds: \cite{Yu_Similar}.
\begin{stat}\label{stat3}$\blacksquare$
Coordinates of the eigenpoints of the dynamical system of the cosmological model $\Mod{M}$ for $H\not\equiv 0$ in the subspace $\mathbb{R}_3\equiv\{H,\Phi,Z\}\subset \mathbb{R }_4$, as well as the eigenvalues of the characteristic matrix, coincide with the coordinates of the eigenpoints and the eigenvalues of the characteristic matrix of the vacuum-field cosmological model. $\blacksquare$
\end{stat}
The coordinates of these six singular points in the subspace $\mathbb{R}_3=\{H,\Phi,Z\}$ of the phase space of the dynamical system are (see \cite{YuKokh_TMF}):
\begin{eqnarray}\label{M_0}
\!\!\!M^\pm_0=\biggl[\pm\sqrt{\frac{\Lambda}{3}},0,0\biggr];
\label{M_pm}
\!\!\!M^\pm_\pm=\biggl[\pm\sqrt{\frac{\Lambda_0}{3}},\pm\frac{m_s}{\sqrt{\alpha}},0\biggr],
\end{eqnarray}
and the non-zero eigenvalues of the characteristic matrix of the system at these singular points are equal to:
\begin{eqnarray}\label{eigen_val_0}
M^\pm_0:& \!\!\!\!
\left|\begin{array}{ll}
\lambda_2=& \pm\sqrt{\displaystyle \frac{\Lambda}{3}},\\[8pt]
\lambda_{3,4}=& \mp\frac{1}{2}\sqrt{3\Lambda} \pm \frac{1}{2}\sqrt{3\Lambda-4m^2_s};\\
\end{array}\right.\\
\label{eigen_val_pm}
M^\pm_\pm:&
\left|\begin{array}{ll}
\lambda_2=& \displaystyle \pm\sqrt{\frac{\Lambda_0}{3}},\\[8pt]
\lambda_{3,4}=& \mp\frac{1}{2}\sqrt{3\Lambda_0} \pm \frac{1}{2}\sqrt{3\Lambda_0+8m^2_s}.\\
\end{array}\right.
\end{eqnarray}
\section{Numerical modeling of the\newline  evolution of spherical\newline  perturbations}
\subsection{Basic functions of the model}
In this section, we will study the $\Mod{M_0}$ model with the following parameter values as a base model
\begin{equation}\label{Par}
\mathbf{P_0} =\bigl[\bigl[1,1,1,0.1\bigr],3\cdot10^{-6}\bigr].
\end{equation}
In this case, a model similar to it with values of parameters \emph{order} of the field theoretical model SU(5) is obtained with a similarity coefficient $k=10^{-5}$:
\begin{eqnarray}\label{ParSU5}
\mathbf{P_{SU(5)}} =\bigl[\bigl[10^{-10},10^{-5},\sqrt{10}\cdot10^{-3},\nonumber\\
\;\;\;\sqrt{10}\cdot10^{-4}\bigr],3\cdot10^{-16}\bigr],\quad (k=10^{-5}),
\end{eqnarray}
and a similar model with \emph{order} parameters of the standard field theoretical model SM - with a similarity coefficient $k=10^{-15}$:
\begin{eqnarray}\label{ParSM}
\mathbf{P_{SM}} =\bigl[\bigl[10^{-30},10^{-15},\sqrt{10}\cdot10^{-8},\nonumber\\
\;\;\;\sqrt{10}\cdot10^{-9}\bigr],3\cdot10^{-36}\bigr],\quad (k=10^{-15}).
\end{eqnarray}
Note that, according to the theorems proved in \cite{Yu_Similar}, in this case the time scales are stretched by $1/k$ times. In addition, note that the values of the initial Fermi momentum used here correspond to nonrelativistic fermions. In the future, we will study the basic model $\Mod{M_0}$ with parameters \eqref{Par} using numerical methods. The reader can easily recalculate the results on the \eqref{ParSU5} and \eqref{ParSM} models independently using the \ref{Similar_table} transformation table\footnote{We deliberately do not present the results of numerical simulations for specific field theoretical models, given the lack of reliable experimental data currently.}. The singular points of the corresponding vacuum-field model, which are also the singular points of the model studied here (see \cite{Yu_Similar}), are\footnote{The coordinates of the singular points are given by the list $[H,\Phi,Z)$].}:
\begin{eqnarray}\label{M_0}
M^\pm_0=\bigl[10^{-3}, 0, 0\bigr];\\
\label{M_pm_real}
M^\pm_\pm=\bigl[[\pm0.2886768667, \pm1, 0]\bigr],
\end{eqnarray}
moreover, the points $M^\pm_\pm$ are saddle points (unstable), and the points $M^\pm_0$ are attracting foci (stable).

Since the value $\Phi=1$ corresponds to an unstable singular point, we consider two types of initial conditions with a scalar potential value near this point:
\begin{eqnarray}\label{I1}{}
\mathbf{I_1}=&[[\Phi_0=0.999, Z_0=0]\textcolor[rgb]{1.00,1.00,1.00}{]},\nonumber\\
&\textcolor[rgb]{1.00,1.00,1.00}{[}[m_0=1,\dot{m}_0=0,q_0=0,\dot{q}_0=0]];\\
\label{I2}{}
\mathbf{I_2}=&[[\Phi_0=1.0001, Z_0=0]\textcolor[rgb]{1.00,1.00,1.00}{]},\nonumber\\
&\textcolor[rgb]{1.00,1.00,1.00}{[}[m_0=1,\dot{m}_0=0,q_0=0,\dot{q}_0=0]].
\end{eqnarray}
In Fig. \ref{Ignatev1} shows graphs of the scale functions of the model $\xi(t)$ and $H(t)$ for these two types of initial conditions.

The initial singularity is reached at time $t_0\approx-6.793$ in the case \eqref{I1} and
$t_0\approx-6.62$ in the case of \eqref{I2}. Note that in the models under study there are two stages of inflationary expansion ($H=H_0=\mathrm{Const}, \Omega=1$).
Although the scale function plots are almost identical for these two types of initial conditions, note that the two models enter the second inflation regime with different expansion rates:
$H_1\approx 1.83\cdot10^{-2}$ and $H_2\approx 2.47\cdot10^{-2}$. In Fig. \ref{Ignatev2} shows the evolution of the scalar field potential and its derivative.
Graphs in Fig. \ref{Ignatev2} demonstrate the transition of the model from the unstable state $\Phi\approx1$ to the stable state $\Phi\to0$, and the transition process is accompanied by potential oscillations that decay with time. The corresponding graphs for the initial conditions \eqref{I1} are practically the same.
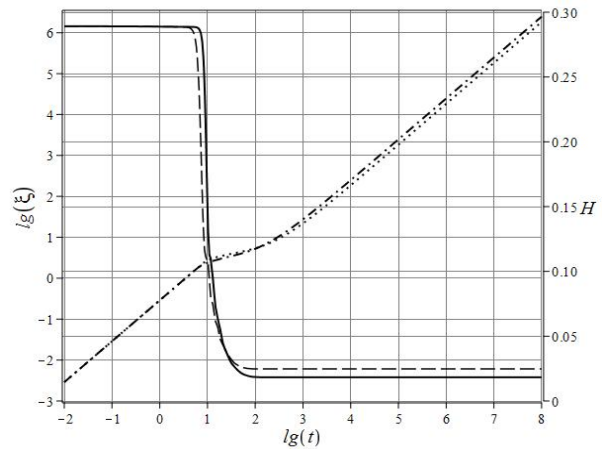
\Fig{Ignatev1.jpg}{6}{\label{Ignatev1}Evolution of scale functions $\xi(t)$ (dashed-dotted line for initial conditions \eqref{I2} and dotted line for \eqref{I1}) and $ H(t)$ (solid line for initial conditions \eqref{I2}) and dotted line for \eqref{I1}) for the model with parameters \eqref{Par}.}
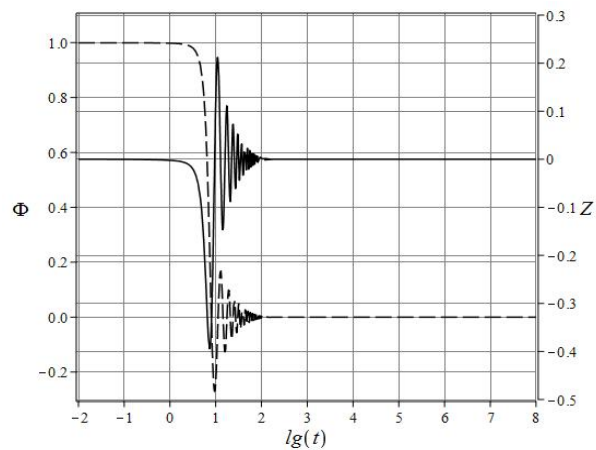
\Fig{Ignatev2.jpg}{6}{\label{Ignatev2}Evolution of the potential $\Phi(t)$ (dashed line) and its derivative $Z(t)$ (solid line) for initial conditions \eqref{I2} ) for a model with parameters \eqref{Par}.}

Further, along with the model of spherical pertur\-ba\-tions \eqref{Par} with initial conditions \eqref{I1} (model $\mathbf{MS_1}$) and \eqref{I2} (model $\mathbf{MS_2}$), we will also consider simplified spherical perturbation models with given values of background functions $[\xi=H_0 t,H=H_0,\Phi=0.025,Z=0]$, $H_0=\sqrt{\Lambda/3}$ (model $\mathbf{M_{mod} }$) and a qualitative model of the evolution of spherical modes of disturbances with wave number $n$ with background functions of the model $\mathbf{MS_1}$, (model $\mathbf{M_{bh}}$), which we used in the first part of the work \cite{Yu_GC_23_No4}\footnote{In this case, we will not take into account the evaporation of black holes, which is significant only at times of the order of several Planck times; for details, see \cite{Yu_GC_23_No4}.}.

\Marking{to the initial conditions\label{mark_inits}}{Note that the initial conditions \eqref{I1} and \eqref{I2} are formulated in such a way that the initial value of the central mass in the perturbation is equal to one Planck mass ($m_0=1$ ), while the initial value of the central charge is assumed to be zero ($q_0=0$). These are, of course, very small values; increasing $m_0$ by $n$ times, we must increase the values $m(t),q(t)$ by the same number of times.}
\subsection{Evolution of mass and charge in spherical perturbations}
In Fig. \ref{Ignatev3} and \ref{Ignatev4} show graphs of the evolution of the central mass $m(t)$ and charge $q(t)$ in the models of spherical perturbations $\mathbf{MS_1}$ and $\mathbf{MS_2}$, respectively.

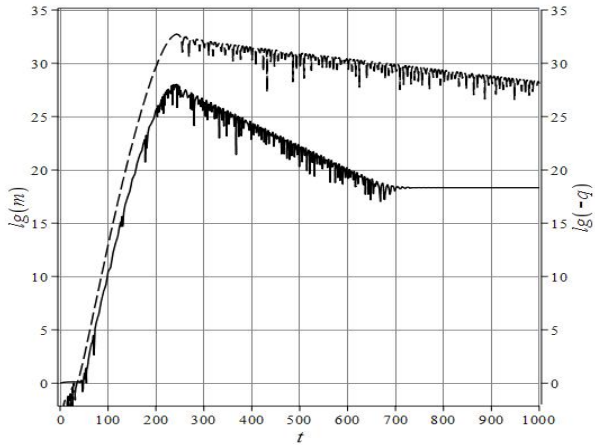
\Fig{Ignatev3.jpg}{6}{\label{Ignatev3}The evolution of the mass $m(t)$ for the model $\mathbf{MS_1}$ is the solid line and the charge $q(t)$ is the dashed line. }

Commenting on the graphs in Fig. \ref{Ignatev3} and \ref{Ignatev4}, note, firstly, that the instability of the state $M^\pm_\pm$ \eqref{M_pm_real}, corresponding to the value of the scalar potential $\Phi=1$ leads to a striking difference in behavior functions $m(t)$ and $q(t)$ (by 25 orders of magnitude!) when the process starts from two very close states, differing from each other in the value of the potential $2\cdot10^{-4}$ from its absolute value. Starting the process from a state below the singular point does not allow reaching the required masses in the process of cosmological evolution.
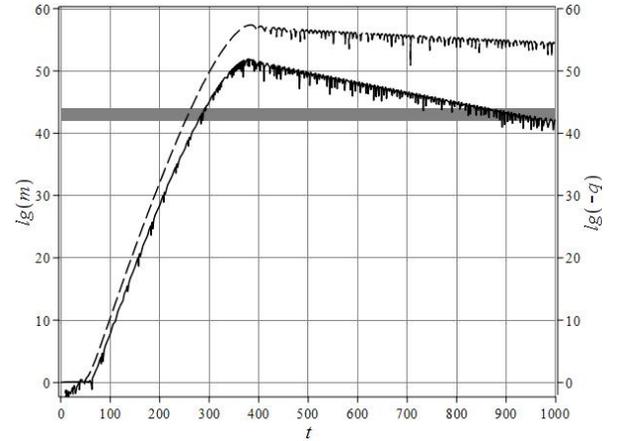
\Fig{Ignatev4.jpg}{6}{\label{Ignatev4}The evolution of the mass $m(t)$ for the model $\mathbf{MS_2}$ is the solid line and the charge $q(t)$ is the dashed line. The gray bar corresponds to the range of mass values required for the formation of SSBH.}
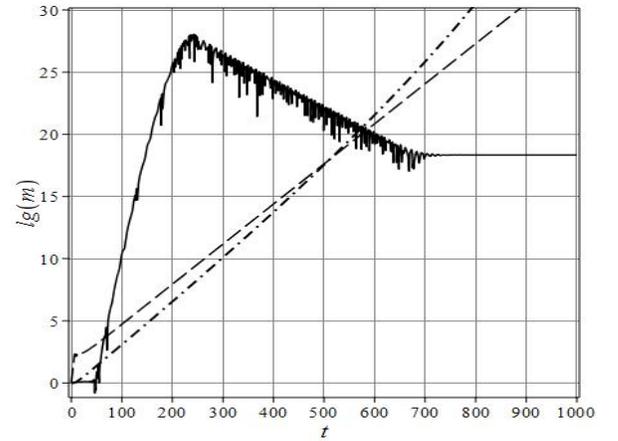
\Fig{Ignatev5.jpg}{6}{\label{Ignatev5}The evolution of mass $m(t)$ for the model $\mathbf{MS_1}$ is a solid line, for the model $\mathbf{M_{mod}}$ -- dash-dotted line and for the model $\mathbf{M_{bh}}$ -- dashed line.}

Secondly, in contrast to the evolution of mass discussed in the first part of the article, the oscillatory nature of this process is observed, apparently caused by damped oscillations of the scalar field during the transition from an unstable state to a stable one. This feature can be easily detected in the graphs \ref{Ignatev5} and \ref{Ignatev6}.

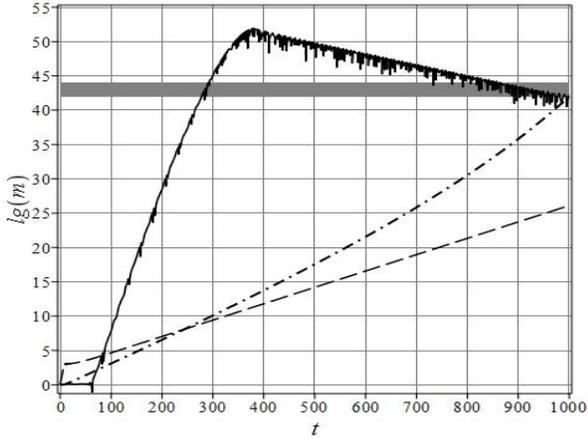
\Fig{Ignatev6.jpg}{6}{\label{Ignatev6}The evolution of mass $m(t)$ for the model $\mathbf{MS_2}$ is a solid line, for the model $\mathbf{M_{mod}}$ -- dash-dotted line and for the model $\mathbf{M_{bh}}$ -- dashed line.}

Thirdly, we note a very important circumstance: in contrast to the models $\mathbf{M_{mod}}$ and $\mathbf{M_{bh}}$, in which the mass $m(t)$ shows unlimited growth of mass, in models of spherical perturbations $\mathbf{MS_1}$ and $\mathbf{MS_2}$ after reaching the maximum value, both the mass and the absolute value of the charge begin to fall. We will consider this feature of the process at the end of the article.

Finally, we note that in the spherical perturbation model the singular mass increases faster than in the $\mathbf{M_{mod}}$ and $\mathbf{M_{bh}}$ models. In Fig. \ref{Ignatev7} shows graphs of the evolution of the Schwarzschild radius of the singular mass $r_g=2m(t)$ and the physical radius $R(t)=a(t)r_0$ for the model $\mathbf{MS_1}$. From this figure it follows that the gravitational collapse of the disturbance ($r_g(t)>R(t)$) occurs in the early stages of cosmological evolution ($t<100$), which was also shown in the first part of the article \cite{Yu_GC_23_No4}.

Let's devote a few lines to explain the rules for using graphs and the Table \ref{Similar_table}. On the graphs Fig. \ref{Ignatev1} -- \ref{Ignatev6} when moving from a model with parameters $\mathbf{P_0}$ to a model $\mathbf{P_{SU(5)}}$ with parameters \eqref{ParSU5} we must stretch time scales by $10^5$ times, while maintaining the values $\xi(t), \Phi)(t)$ and compressing the scale of values $Z(t)$ by $10^5$ times, and in the case of the model $\ mathbf{P_{SM}}$ -- $10^{15}$ times. On the graphs Fig. \ref{Ignatev3} -- \ref{Ignatev6} when saving the initial conditions \eqref{I2} in the model with parameters $\mathbf{P_{SU(5)}}$ \eqref{ParSU5} we must save the value $m( t)$ and $q(t)$ (see Note \ref{mark_table}), but at the same time the time scales on the graph will be stretched by $10^5$ times, in the case of a model with parameters $\mathbf{P_{SM}} $ \eqref{ParSM} time scales need to be stretched $10^{15}$ times.

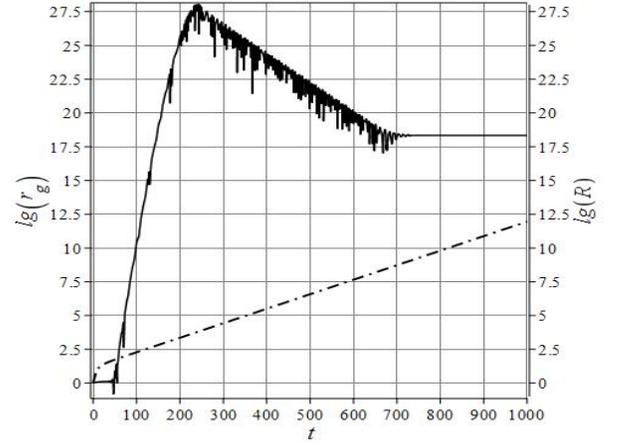
\Fig{Ignatev7.jpg}{6}{\label{Ignatev7}Evolution of the Schwarzschild radius of the singular mass $r_g=2m(t)$ (solid line) and the physical radius $R(t)=a(t)r_0$ (dashed -dotted line) for model $\mathbf{MS_1}$.}

\subsection{The relationship between the linear\newline model of the evolution of spherical\newline disturbances and the complete\newline nonlinear picture}
Let us note once again that the quantities $m(t)$ and $q(t)$ are singular mass and charge of the black hole according to \eqref{nu=1-2} -- \eqref{phi=1-2} and \eqref {shvarc}.As is known from the standard theory of spherical gravitational
collapse (see, for example, \cite{Land_Field}) in the process of gravitational collapse, the mass of a black hole cannot decrease. How then can we understand the results presented in the graphs in Fig. \ref{Ignatev3} -- \ref{Ignatev6}, on which the mass $m(t)$, after reaching the maximum value, begins to slowly fall. Here we must remember that our model is a linear model of the evolution of spherical disturbances, which, in principle, can allow such behavior, apparently caused by a critical increase in pressure near the linear singularity.

A nonlinear model of the spherical collapse of scalarly charged fermions should lead to a fixation of the black hole mass when it reaches its maximum. Therefore, it is likely that the maximum of the function $m(t)$, achieved during the evolution of spherical perturbations, should be the final mass of the SSBH. This question, of course, requires additional and very complex research.

Thus, the drop in the black hole mass function after reaching its maximum value is a consequence of the violation of the linearity of the black hole model. It must be assumed that the oscillations of the mass and charge functions, which can be observed on the presented graphs, which are a consequence of small oscillations of the scalar field, in the nonlinear picture should be replaced by oscillations of the first derivatives of these quantities.
\section*{Conclusion}
Summing up the intermediate results of this part of the article, we note, firstly, on the basis of a more adequate theory of spherical perturbations, it was possible to confirm the fundamental possibility of achieving, during the development of scalar-gravitational instability of spherical perturbations, singular masses of the order of magnitude necessary for the formation of SSBH \eqref{M_nc}. Secondly, we managed, in principle, to solve the problem outlined at the beginning of the article - the problem of stopping the unlimited growth of the disturbance mass. We will devote the third and final part of the article to solving the second problem identified at the beginning of the article: \\
\emph{What does the large-scale structure of the Universe become after the completion of this process, what is the fate of the matter that fell into the sphere of influence of SSBH?}

\noindent \textbf{Acknowledgements}\\
The author is grateful to the participants of the seminar of the Department of Relativity and Gravity at Kazan University for a useful discussion of some aspects of the work. The Author is especially grateful to professors S.V. Sushkov and A.B. Balakin.

\noindent \textbf{Founding}\\
The work was carried out using subsidies allocated as part of state support for the Kazan (Volga Region) Federal University in order to increase its competitiveness among the world's leading scientific and educational centers.

\setcounter{section}{0}
\setcounter{equation}{0}
\setcounter{figure}{0}


\end{document}